\documentclass[conference]{IEEEtran}
\IEEEoverridecommandlockouts
\usepackage{balance}
\usepackage{cite}
\usepackage{amsmath,amssymb,amsfonts}
\usepackage{algorithmic}
\usepackage{graphicx}
\usepackage{textcomp}
\usepackage{xcolor}
\usepackage{comment}
\usepackage[export]{adjustbox}
\usepackage{multirow}

\def\BibTeX{{\rm B\kern-.05em{\sc i\kern-.025em b}\kern-.08em
    T\kern-.1667em\lower.7ex\hbox{E}\kern-.125emX}}
\begin{document}

\title{LLMs with User-defined Prompts as Generic Data Operators for Reliable Data Processing
}

\author{\IEEEauthorblockN{Luyi Ma$^*$, Nikhil Thakurdesai$^*$, Jiao Chen$^*$, \\ Jianpeng Xu, Evren Korpeoglu, Sushant Kumar, Kannan Achan}

\IEEEauthorblockA{\textit{Personalization Team} \\
\textit{Walmart Global Tech}\\
Sunnyvale, CA, USA \\
\{luyi.ma, nikhil.thakurdesai, jiao.chen0,\\ jianpeng.xu, ekorpeoglu, sushant.kumar, kannan.achan\}@walmart.com}
}

\maketitle

\def\thefootnote{*}\footnotetext{Equal Contribution}\def\thefootnote{\arabic{footnote}}

\begin{abstract}
Data processing is one of the fundamental steps in machine learning pipelines to ensure data quality. 
Majority of the applications consider the user-defined function (UDF) design pattern for data processing in databases. 
Although the UDF design pattern introduces flexibility, reusability and scalability, the increasing demand on machine learning pipelines brings three new challenges to this design pattern -- not low-code, not dependency-free and not knowledge-aware. 
To address these challenges, we propose a new design pattern that large language models (LLMs) could work as a generic data operator (LLM-GDO) for reliable data cleansing, transformation and modeling with their human-compatible performance.  
In the LLM-GDO design pattern, user-defined prompts (UDPs) are used to represent the data processing logic rather than implementations with a specific programming language. 
LLMs can be centrally maintained so users don't have to manage the dependencies at the run-time. 
Fine-tuning LLMs with domain-specific data could enhance the performance on the domain-specific tasks which makes data processing knowledge-aware. 
We illustrate these advantages with examples in different data processing tasks. 
Furthermore, we summarize the challenges and opportunities introduced by LLMs to provide a complete view of this design pattern for more discussions. 
\end{abstract}


\section{Introduction}

Machine learning (ML) powers numerous data-driven applications for varied types of use cases.
A typical machine learning pipeline consists of data processing, feature engineering, model selection, model training, hyper-parameters tuning, evaluation, testing, and serving \cite{zaharia2018accelerating}. Many of these steps not only require high-quality data to ensure the machine learning applications will perform as expected, but also prefer huge volume of data to support the training \cite{deng2009imagenet}. 

However, most of the reliable datasets were generated by human annotations. This process couldn't scale up well as it is both expensive and time-consuming, limiting its further applications.  
Moreover, with the increasing parameters and complexity of machine learning models, huge volume of high-quality data are required for model training.
The data processing tasks need to support the growing demand of effective data cleansing, transformation and modeling. Following the definition of a typical ETL (Extract, Transform, Load) process in data warehousing, our main focus is on the transform process. 

\begin{figure}
    \centering
    \includegraphics[width=\linewidth]{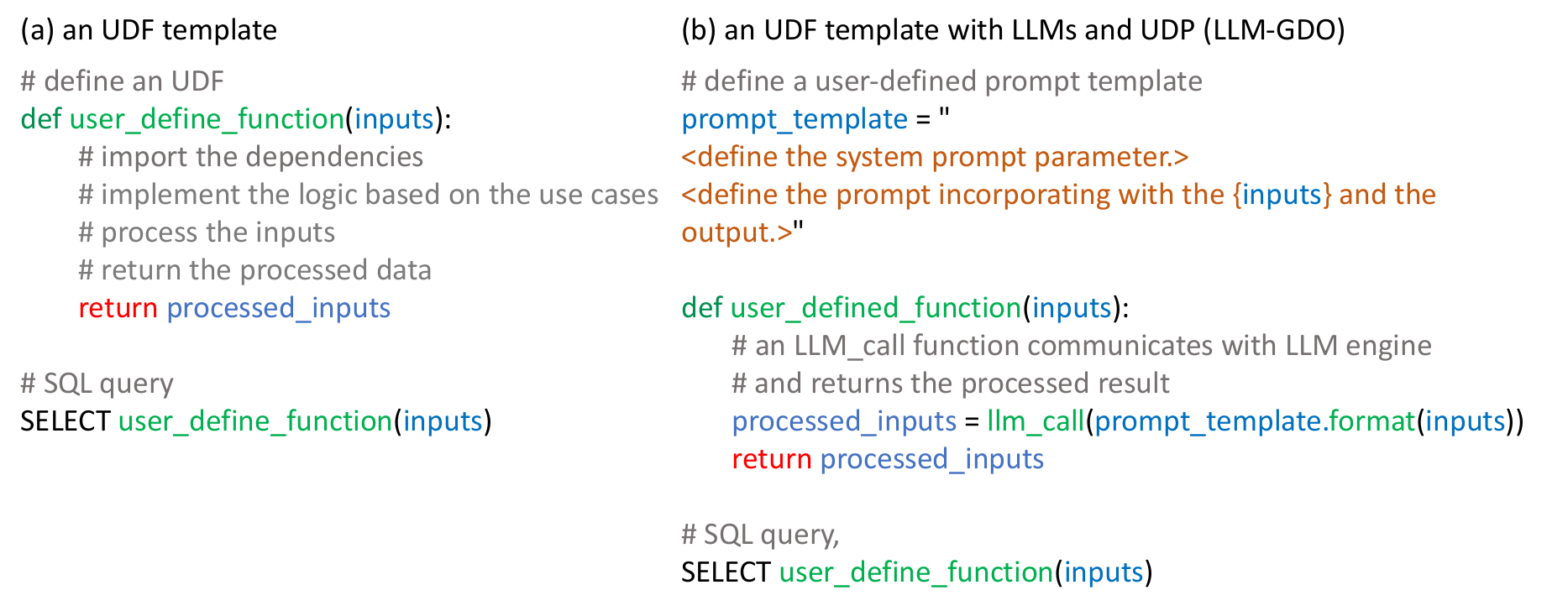}
    \caption{(a) UDF design pattern and (b) LLM-GDO design pattern.}
    \label{fig:template}
\end{figure}

To support this growing demand of data transformation, user-defined functions (UDFs) are commonly used to clean, transform and model data in a data warehouse or a data lake \cite{nambiar2022overview}. 
A typical UDF template is shown in Figure \ref{fig:template}-(a) with a pythonic way of coding. 
Within the UDF, a user could import the run-time dependencies, implement the logic and process the input data. 
When applying the UDF on a database, following the classic narrow transformation setting in Spark, the UDF will be applied to each row of the data and the processed row will be stored \footnote{https://www.databricks.com/glossary/what-are-transformations}. 
The UDF design pattern introduces three advantages in large-scale data processing. 
First, it provides the flexibility for users to implement their own data processing logic that are not supported by built-in functions. 
Second, it abstracts the functionalities for better understanding, debugging and reusability (modular programming).
Third, it can be easily scaled up by big data processing engine like Spark \cite{zaharia2016apache}.
With the above advantages, a user could implement an UDF with a programming language supported by the system (e.g., Python in a PySpark cluster built on a Hadoop file system), and apply the logic in parallel over all the records. 
However, this design pattern also meets increasing challenges.
(1) \textbf{Not low-code or zero-code}: it requires the users to have substantial programming skills and experiences.
(2) \textbf{Not dependency-free}: it could require a complicated run-time environment for different UDFs. Managing the dependencies is difficult in both development and deployment. For example, if the sets of run-time dependencies for two UDFs have no overlap, we need two separated pipeline to manage the dependencies. 
(3) \textbf{Not knowledge-aware}: it is difficult to natively incorporate prior knowledge for the data processing logics into current implementations of UDFs. 
Prior knowledge is usually task-specific. 
For instance, e-commerce item category classification requires a strong domain knowledge to identify the item attributes.
It is difficult to implant this knowledge into UDFs deterministically for item classification due to the enormous combinations of item attributes.

Recently, artificial intelligence (AI) development has been gaining a promising progress in the past years with the emergence of Large Language Models (LLMs). 
LLMs, such as Llama2 \cite{touvron2023llama} and GPT-4 \cite{openai2023gpt4}, show their effectiveness in solving a wide range of downstream tasks (e.g., question answering, multi-step reasoning etc.) due to the emergent abilities \cite{wei2022emergent}, which reduces the gap between natural language and programming. 
With well-designed prompts, a user without sufficient experience in data processing can easily employ an LLM to extract the aspects of products, which usually requires the knowledge of e-commerce domain experts \cite{maragheh2023llm}. This learning ability, which only relies on natural language instructions and input-output examples, and without optimizing any parameters, is called in-context learning \cite{brown2020language}. 
This in-context learning ability advances LLMs to understand few-shots even zero-shot learning tasks, for instance, classifying the tabular\cite{hegselmann2023tabllm} and anomaly detection in system logs \cite{qi2023loggpt}.
Similar to other pre-trained models, LLMs' performance could be further improved by fine-tuning with different techniques, for example, LoRA \cite{hu2021lora} and QLora \cite{dettmers2023qlora} can efficiently fine-tune LLMs by optimizing the rank decomposition matrices of the dense layers in a neural network.
The fine-tuned LLMs could achieve human-compatible results in many tasks \cite{openai2023gpt4}\cite{chen2023knowledge}, greatly reducing the human effort in labeling and annotation.

In our paper, we address the limitations in the current UDF-based data processing practice and summarize a new design pattern for data processing involving LLMs and user-defined prompts (\textbf{UDPs}) to balance flexibility and human-level accuracy.
We propose a design pattern that LLMs with UDPs could work as \textit{G}eneric \textit{D}ata \textit{O}perators (\textbf{LLM-GDOs}) for data cleansing, transformation and modeling. 
To visualize this design pattern, we show an example of LLM-GDO in Figure \ref{fig:template}-(b).
In LLM-GDO design, we simplify the UDF with two changes.
First, instead of defining a programming-language-based UDF, a user could define a prompt (or a prompt template for better representation) `{\color{blue}user\_defined\_prompt}' to describe the data processing logic (low-code and zero-code). When the data distribution is changed, instead of updating the processing code, UDPs can update the data processing logic easily with modifying the instructions and examples in prompts.
Unlike an UDF which requires run-time dependencies to support the execution, an LLM (pre-trained or fine-tuned) could work as a compiler for the prompt and execute the request independently (dependency free). 
As we use the same LLM with a proper version control, we can align the offline development and online serving.
When processing data, the database talks with the remote LLM resource via LLM gateways (e.g., APIs or Agents) behind which LLMs are maintained to process the requests. 
We abstract this function as `{\color{green}llm\_call}' and the implementation details could be done by platforms and encapsulated well from users. 
By fine-tuning LLMs, we can seamlessly introduce the domain-specific knowledge into LLMs with a small dataset and enhance their performance on these tasks (knowledge-aware). 
Although LLMs are versatile, they also have limitations which we should keep improving.
To provide a complete view of this design pattern, We foresee the challenges in this design pattern.

Our contributions are summarized as follows:
\begin{itemize}
    \item we introduce the design pattern LLM-GDO in the big data setting for ML pipelines.
    \item we summarize the potential applications with LLM-GDOs. 
    \item we discuss the challenges and opportunities in LLM-GDO.
\end{itemize}

The paper is structured as follows. 
We will introduce the key concepts in Section \ref{preliminaries} and present a comprehensive comparison between UDFs and LLM-GDOs in Section \ref{method}, followed with challenges and opportunities in Section \ref{eval}. Finally we conclude our paper in Section \ref{conclusion}.

\section{Preliminaries} \label{preliminaries}
\subsubsection{Narrow Transformations and Wide Transformations}
\begin{figure}
    \centering
    \includegraphics[width=0.8\linewidth]{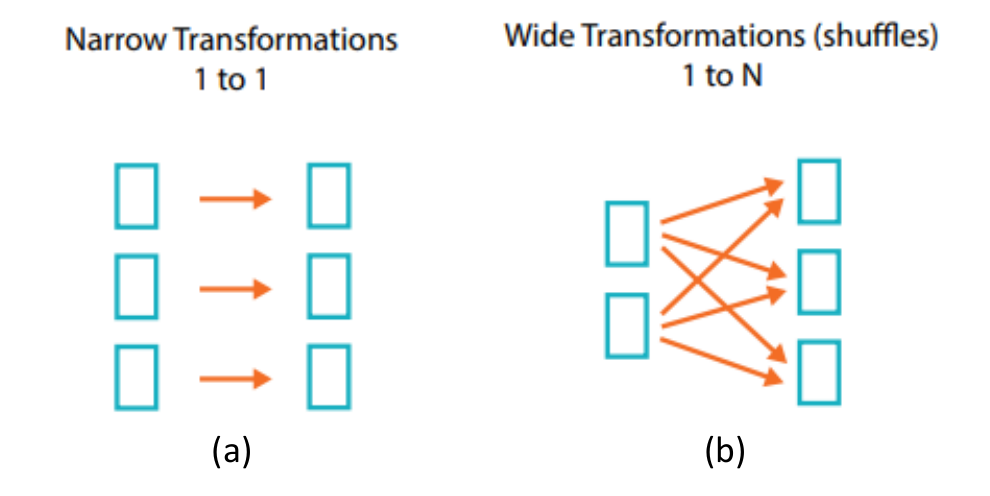}
    \caption{(a) Narrow Transformations and (b) Wide Transformations}
    \label{fig:narrow-wide-transformations}
\end{figure}

Data transformations are instructions of modifying the rows of database. 
In Spark, depending on the dependencies between data points, data transformations could be grouped into narrow transformations (Figure \ref{fig:narrow-wide-transformations}-(a)) and wide transformations (Figure \ref{fig:narrow-wide-transformations}-(b)).
Typically, the output of a narrow transformation operation depends on only one input data (no data shuffling), while output of a wide transformation depends on multiple input rows (with data shuffling).
In our paper, we focus on the narrow transformations as they are dominant in many early-stage data processing steps to improve the data quality. 

\subsubsection{LLMs Function calling and Fine-tuning}
There are many ways of accessing LLMs, e.g. OpenAI API \footnote{https://platform.openai.com/docs/api-reference} provides one of the popular solutions. 
In our paper, we assume that the LLM's gateways are predominantly maintained by the platforms (remote or local). 
In Figure \ref{fig:template}-(b), we abstract the LLM function calling by the `{\color{green}llm\_call}' function. 
This function takes the formatted UDP and returns the processed output. 
LLMs could be fine-tuned by a small sample of high-quality data. 
As new data flows into the database, the platform could seamlessly conduct fine-tuning of LLMs by extracting a small sample of high-quality data. 
Although there are still many challenges with LLM fine-tuning, it is beyond the scope of this discussion.

\section{Methodology} \label{method}
In this section, we present a list of tasks where LLM-GDOs could improve UDFs. 
Note that LLMs are still evolving so we mainly focus on the design pattern in this section to address the connection between UDFs and LLM-GDOs. 
We will provide comprehensive discussion about the challenges of current LLM-GDO design in Section \ref{eval}.
For brevity, we reuse the definition of `{\color{green}user\_defined\_function}' in Figure \ref{fig:template}-(b) in the following case studies. All the example LLM outputs in this paper have been generated using Chat-GPT 3.5 \cite{chatgpt35turbo}.

\begin{figure}
    \centering
    \includegraphics[width=0.9\linewidth]{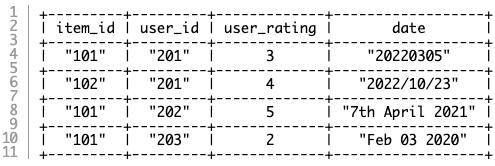}
    \caption{A sample table `item\_rating' with columns `item\_id', `user\_id', `user\_rating' and `date'.}
    \label{fig:sample_table}
\end{figure}

\subsection{Data Cleansing and Transformation}
Data cleansing and transformation are crucial steps to ensure the performance of a ML pipeline. 
To compare UDFs and LLM-GDOs in data cleansing and transformation, we consider a sample table `item\_rating' defined in Figure \ref{fig:sample_table} with the following three tasks. They illustrate the low-code feature and dependency-free feature of LLM-GDO.

\subsubsection{Data Structural Consistency}
\begin{figure}
    \centering
    \includegraphics[width=0.9\linewidth]{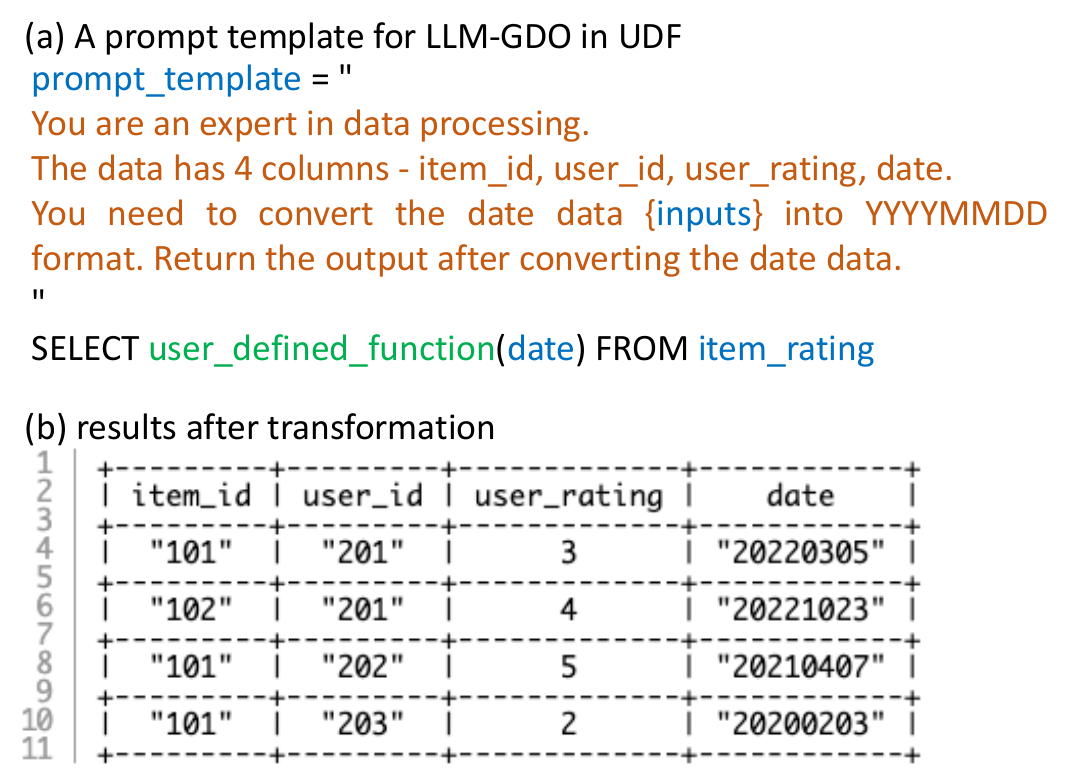}
    \caption{LLM-GDO-based data transformation for data structural consistency.}
    \label{fig:data_structure}
\end{figure}

Data structural consistency is usually one of the initial steps. 
It helps to structuralize the data to improve the downstream transformations. 
In the `item\_rating' table (Figure \ref{fig:sample_table}), the `date' column contains date strings in different formats. 
Figure \ref{fig:data_structure} shows an example to structuralization the date data. 
With LLM-GDO in UDF, we can define the output format (YYYYMMDD) in the prompt and let LLMs handle the data processing (Figure \ref{fig:data_structure}-(b)).
Traditional UDFs can also complete this structuralization but it usually requires either enumeration of the date format or leverage different packages. 

\subsubsection{Data Type Conversion}
\begin{figure}
    \centering
    \includegraphics[width=0.9\linewidth]{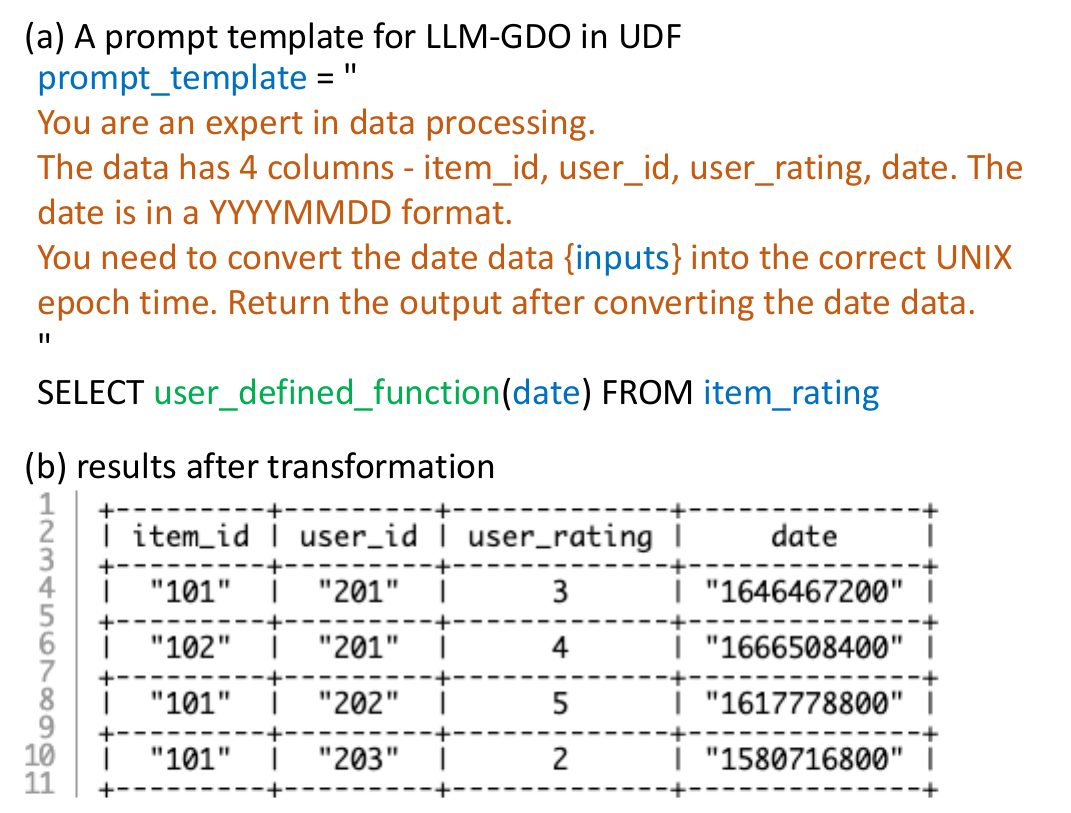}
    \caption{LLM-GDO-baesd data transformation for data type conversion.}
    \label{fig:data_convert}
\end{figure}

After cleaning the data structure with better structural consistency, we can convert the data from one type to another. 
Figure \ref{fig:data_convert} presents an example of UNIX epoch time conversion from the given date data. 
Again, the LLM-GDO completes this data transformation with the instruction in the prompt, while the traditional UDF users need to understand the definition of UNIX epoch time for implementation or know the right packages to call. 

\subsubsection{Data Standardization}
\begin{figure}
    \centering
    \includegraphics[width=0.9\linewidth]{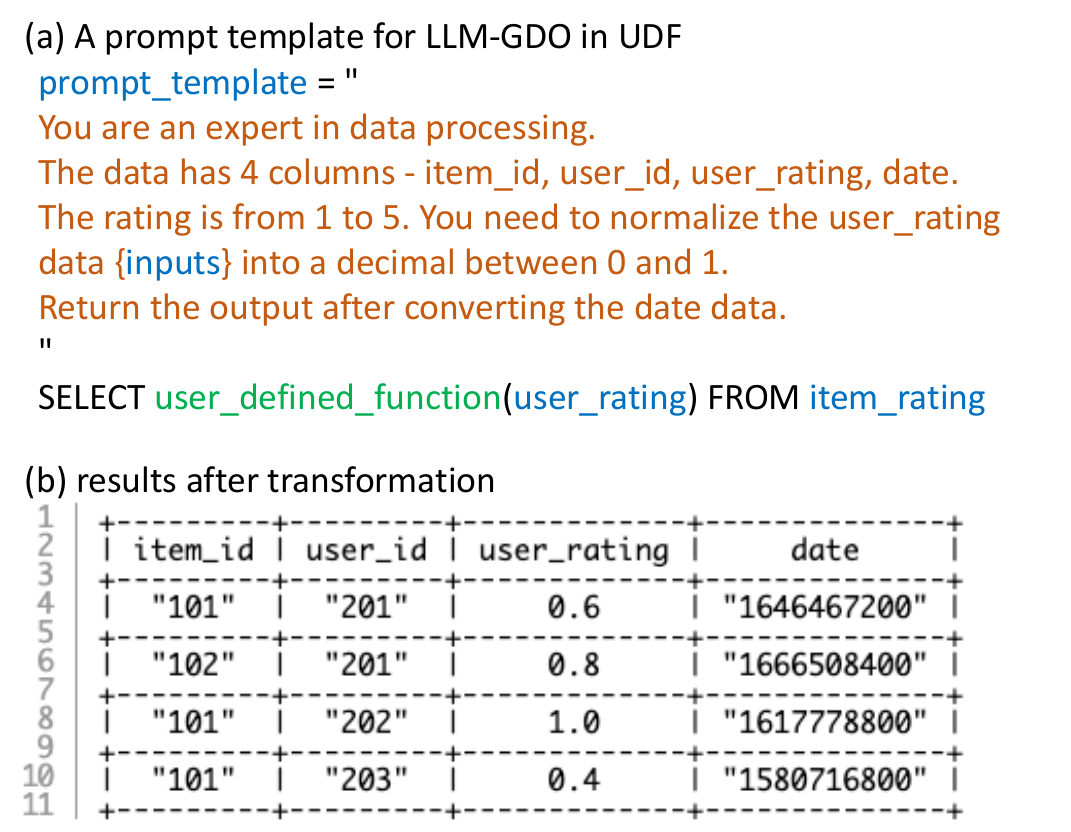}
    \caption{LLM-GDO-baesd data transformation for standardization.}
    \label{fig:data_strandarization}
\end{figure}

Another important task is normalizing the numeric data. 
In many machine learning pipelines, normalization of numeric values increases the stability of the model training. 
In Figure \ref{fig:data_strandarization}, we present an example wehre LLM-GDO is employed to normalize the `user\_rating' column by providing the rating range . 


\subsection{Data Modeling}
\begin{figure}
    \centering
    \includegraphics[width=\linewidth]{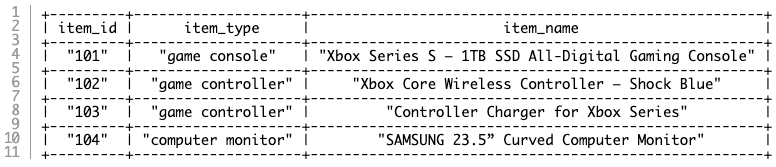}
    \caption{A sample table `item\_information' with columns 'item\_id', `item\_types' and `item\_name'.}
    \label{fig:sample_table2}
\end{figure}

In addition to data cleansing and transformation, many feature engineering steps require machine learning models to conduct the basic reasoning and generate high-quality features. 
For example, in many NLP tasks, word annotations and tagging are important features for modeling \cite{owoputi2013improved} \cite{hu2017wordsup}.
These features usually require a dedicated machine learning pipeline to extract the features, which are inconvenient to maintain without a proper support from ML operations. 
A user needs to understand the machine learning pipeline in detail to run an UDF to employ these complicated models, managing the dependencies and another layer of data processing. 
LLM-GDOs, however, can keep the same convenience with a uniform practice. 
We consider a new sample table `item\_information' defined in Figure \ref{fig:sample_table2} for item information. 
We highlight this advantage with the following two tasks. 

\subsubsection{Reasoning} 
\begin{figure}
    \centering
    \includegraphics[width=\linewidth]{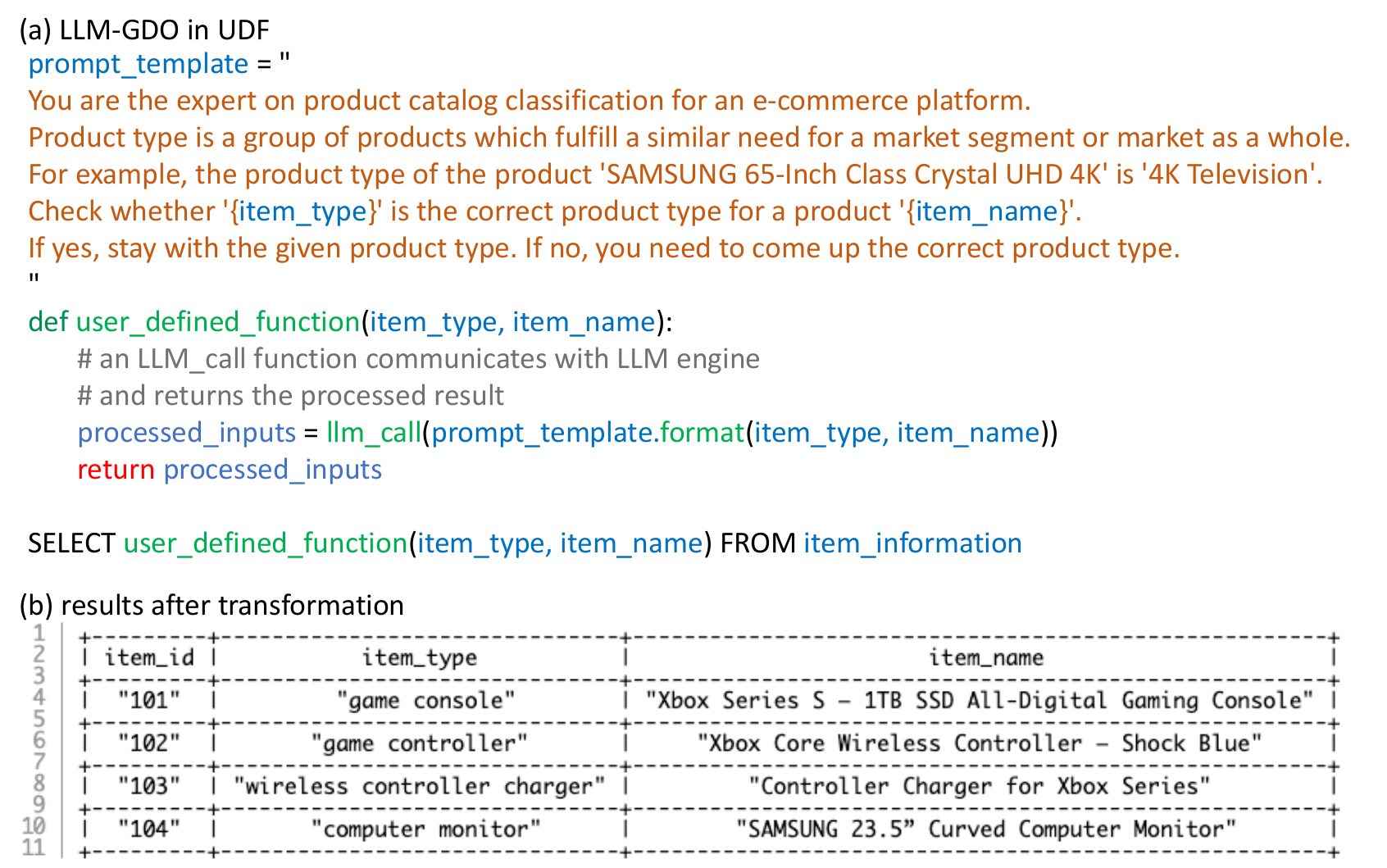}
    \caption{An LLM-GDO application for data reasoning and anomaly detection on item type classification.}
    \label{fig:anomaly_detection}
\end{figure}

In this task, we need to parse each row of data and conduct reasoning (e.g., classification) to get the output. 
One of the common use cases is to detect the anomaly value in the database. 
For example, in many system log databases, the system error messages are grouped under several categories by their severity levels or the types of events. 
Another example is that items are grouped into product types in e-commerce for item display. 
The classification could be invalid but the anomaly is hard to find due to the huge data volume. 
Anomaly detection could be done by separate machine learning pipelines \cite{pang2021deep} but integrating them into database is challenging. 
First, these machine learning pipelines have different dependencies and require domain knowledge. 
Second, as the underlying data changes, the models might not be updated without a proper orchestration system for model retraining. 
LLM-GDO can conduct reasoning to detect the anomaly in this case. 
Figure \ref{fig:anomaly_detection} shows an example of LLM-GDO for anomaly detection in the `item\_information' table. 
We can see that LLM-GDO can easily detect the wrong item type for the `103' item in Figure \ref{fig:sample_table2} and correct it in Figure \ref{fig:anomaly_detection}-(b). 
In this prompt, we can leverage the in-context learning of LLM by designing a prompt \cite{hegselmann2023tabllm}.
Compared with the traditional UDF, users have to define the complex heuristic or different versions of deep learning models to complete the task. 
The LLM-GDO simplify the development of the logic and maintenance because LLMs at the back-end could be managed in a centralized way which powers the tasks without disclosing the details to users. 

\subsubsection{Embedding Database}
Embedding is another important use case for data processing. 
With LLM-GDO, we can also generate high-quality embedding based on the `item\_name' column \footnote{https://platform.openai.com/docs/guides/embeddings}. 
Due to the limited space, we skip the prompt example and the output embeddings. 
However, with this approach, we can timely update the embedding for downstream modeling.

\section{Discussion} \label{eval}
Despite the remarkable human-compatible and programmable performance of LLMs in data transformation, it is crucial to highlight the numerous opportunities and challenges present in the LLM-GDO design pattern. This will help to shed light on potential future research directions and enhance our understanding of this evolving field.

\paragraph{LLM inference and Scalability}

Although LLMs demonstrate greater scalability compared to traditional UDFs, they demand significantly more computational resources. This can be attributed to the massive number of parameters in LLMs, which require extensive computational resources for model inference. For instance, a GPT-3-sized LLM requires the use of eight 80G A100 GPUs for inference\footnote{https://blogs.oracle.com/research/post/oracle-first-to-finetune-gpt3-sized-ai-models-with-nvidia-a100-gpu}. Moreover, current commercial LLMs, such as GPT-3.5 and PaLM2, have limited speed in response.  Nonetheless, research on LLM knowledge distillation\cite{hsieh2023distilling} and compression\cite{zhu2023survey} are promising avenues for reducing LLM size and computational demands, thereby making them more cost-effective, efficient, and scalable.

\paragraph{LLM Hallucination}
Hallucination means a LLM makes up contents that do not exist in real world or do not satisfy the requirements in the prompt. While hallucination can be a feature for many creative use cases, it is usually an obstacle for data processing \cite{zhang2023language} . 
For instance, when the desired output data format is tab-separated values (TSV), it is crucial to avoid inconsistencies, such as the inclusion of alternative separators, particularly when employing an LLM within a production data engineering pipeline.


Various methods have been proposed to mitigate and quantify hallucinations in LLMs, including reducing the model's temperature, providing more context in the prompt, employing a chain-of-thought approach \cite{zhang2022automatic}, ensuring self-consistency \cite{huang2022large}, or specifying more precise formatting requirements within the prompt. Despite these efforts, the effective detection and prevention of hallucinations in LLMs during data processing remains a formidable challenge.





\paragraph{LLM Unit Testing and Evaluation}
Traditional UDFs are typically deterministic, allowing for the development of unit tests to assess their correctness. 
In contrast, LLMs employ a generative approach to produce results based on probability \cite{chang2023survey}, making it more challenging to conduct universal testing.
Recently, some teams have leveraged the larger LLMs to generate the unit test cases (e.g., expected outputs) for the prospective LLMs \footnote{https://www.anyscale.com/blog/fine-tuning-llama-2-a-comprehensive-case-study-for-tailoring-models-to-unique-applications}.
We believe LLM unit testing will attract more attentions in LLM development and application. 



\paragraph{LLM Privacy}
Fine-tuning an industry-level LLM for data processing could depend on data from different departments, involving data transition and sharing. 
For example, when processing data from a social app, customers' profile and preference are highly sensitive. 
Data sharing will increase the probability of data leaks and violate the data privacy policy \cite{290847}. 
Recent works in federated learning \cite{fan2023fate} address the privacy issues in LLMs. 
We expect more research interests on LLM privacy issues.
\section{Conclusion} \label{conclusion}
In this research paper, we propose a novel design pattern, LLM-GDO, aimed at improving the efficiency and reliability of data transformation. LLM-GDO offers the benefits of utilizing low-code and dependency-free implementations for knowledge-aware data processing. However, it also encounters challenges stemming from LLMs. We examine these challenges and opportunities, and provide an in-depth perspective on the LLM-GDO design pattern.

\bibliographystyle{IEEEtran}
\balance
\bibliography{main} 

\end{document}